\documentclass[oneside,11pt]{article}
\usepackage{amsmath}
\usepackage{amsfonts}
\usepackage{amssymb}
\usepackage{graphicx}

\begin{document}

\title{\textbf{\LARGE Friedmann-Robertson-Walker Models with Late-Time
Acceleration}}
\author{{\large Abdussattar}$^{1}${\large \ and S R Prajapati}$^{2}$ \\
%EndAName
Department of Mathematics, Faculty of Science, \\
Banaras Hindu University, Varanasi -221005, India \\
Email: asattar@bhu.ac.in$^{1}$, shibesh.math@gmail.com$^{2}$}
\date{}
\maketitle

\begin{abstract}
In order to account for the observed cosmic acceleration, a modification of
the ansatz for the variation of density in Friedman-Robertson-Walker (FRW)
models given by Islam is proposed. The modified ansatz leads to an equation
of state which corresponds to that of a variable Chaplygin gas, which in the
course of evolution reduces to that of a modified generalized Chaplygin gas
(MGCG) and a Chaplygin gas (CG), exhibiting late-time acceleration.
\end{abstract}

We consider the homogeneous and isotropic Robertson-Walker space-time 
\begin{equation}
ds^{2}=-dt^{2}+R^{2}(t)\left[ \frac{dr^{2}}{1-kr^{2}}+r^{2}(d\theta
^{2}+\sin ^{2}\theta ~d\phi ^{2})\right]  \label{IC1}
\end{equation}%
described by its scale factor $R(t)$ and curvature parameter $k=0,$ $\pm 1$.
The universe is assumed to be filled with a distribution of matter
represented by the energy-momentum tensor of a perfect fluid given by 
\begin{equation}
T_{ij}=\left( \rho +p\right) U_{i}U_{j}+p~g_{ij}  \label{IC2}
\end{equation}%
where $\rho $ is the energy density of the cosmic matter and $p$ is its
pressure. The Einstein field equations 
\begin{equation}
R_{ij}-\frac{1}{2}R_{k}^{k}g_{ij}=-8\pi GT_{ij}  \label{IC3}
\end{equation}%
for the space-time (\ref{IC1}) yield the following two independent equations 
\begin{equation}
8\pi G\rho =3\frac{\dot{R}^{2}}{R^{2}}+3\frac{k}{R^{2}}  \label{IC4}
\end{equation}

\begin{equation}
8\pi Gp=-2\frac{\ddot{R}}{R}-\frac{\dot{R}^{2}}{R^{2}}-\frac{k}{R^{2}}
\label{IC5}
\end{equation}%
from which we obtain the conservation equation

\begin{equation}
\frac{d}{dR}\left( \rho R^{3}\right) =-3pR^{2}  \label{IC6}
\end{equation}%
and the Raychaudhuri equation

\begin{equation}
8\pi G(\rho +3p)=-6\frac{\ddot{R}}{R}  \label{IC6A}
\end{equation}

In the past few years, there has been spurt activity in discovering the
models of the universe in which the expansion is accelerating, fueled by
some self-interacting smooth unclustered fluid with high negative pressure
collectively known as dark energy\textit{\ } \cite{DE1i}-\cite{DE1iii}.
These models are mainly motivated by the cosmological observations of the
supernovae of type Ia \cite{snei}-\cite{sneiv}. Dark energy is also
supported by other observations, for example, the anisotropy measurements of
the cosmic microwave background radiation \cite{cmbi}-\cite{cmbii} and the
observations of the baryon acoustic oscillations \cite{baoi}-\cite{baoii}.

Dark energy can be represented by a large-scale scalar field $\phi $
dominated either by potential energy or nearly constant potential energy.
Such a matter will also have its energy-stress tensor in the form $%
T_{ij}^{DE}=(\rho _{\phi }+p_{\phi })U_{i}U_{j}+p_{\phi }~g_{ij}$ and its
equation of state in the form $p_{\phi }=w_{\phi }\rho _{\phi }$, where $%
w_{\phi }$ is a function of time in general. A large class of scalar field
dark energy cosmological models have been proposed in recent years,
including cosmological constant $\Lambda $ for which $w_{\phi }$ reduces to
the value -1 (potential energy dominated scalar field) \cite{VISH},
quintessence \cite{DE2i}-\cite{DE2iv}, K-essence \cite{DE3}, tachyon \cite%
{DE4i}-\cite{DE4ii}, Phantom \cite{DE5i}-\cite{DE5ii}, ghost condensate \cite%
{DE6i}-\cite{DE6ii}, quintom \cite{DE7i}-\cite{DE7iv} and spintessence \cite%
{DE8}. Scalar fields are not the only possibility for the dark energy but
there are some alternatives also. Cosmic acceleration can also be accounted
by invoking inhomogeneity \cite{DE9}, \cite{DE10}. It can also be carried
out by using some perfect fluid but obeying "the exotic" equation of state,
the so-called Chaplygin gas \cite{DE11}, \cite{DE12}. Chaplygin gas (CG) is
a peculiar perfect fluid characterized by the equation of state $p=-\frac{A}{%
\rho }$ ($A$ is a positive constant).\ Chaplygin introduced this equation of
state \cite{cha1} as a suitable mathematical approximation for calculating
the lifting force on a wing of an airplane in aerodynamics. The same model
was rediscovered later in the same context \cite{cha2}-\cite{cha3}. The
negative pressure following from the Chaplygin equation of state could also
be used for the description of certain effects in deformable solids \cite%
{cha4}, of stripe states in the context of the quantum Hall effect and of
other phenomena. The Chaplygin gas emerges as an effective fluid associated
with d-branes \cite{DE13i}-\cite{DE13ii} and can also be derived from
Born-Infeld type Lagrangians \cite{DE12}, \cite{DE14}. One of its most
remarkable property is that it describes a transition from a decelerated
cosmological expansion to a stage of cosmic acceleration.

Islam \cite{ISLAM} has made the ansatz, in which the mass-energy density $%
\rho $ in Friedmann-Robertson-Walker (FRW) models is given as a function of $%
R$ as

\begin{equation}
\rho =\frac{A}{R^{4}}\left( R^{2}+b\right) ^{1/2}  \label{IC8}
\end{equation}%
where $A$ and $b$ are positive constants and obtained exact solutions
connecting radiation and matter eras for all the three cases of FRW models.
For small $R$, the function $\rho $ behaves like $R^{-4}$, while for large $%
R $, it behaves like $R^{-3}$, (the cases of pure radiation and zero
pressure of standard models). All the FRW models based on the ansatz (\ref%
{IC8}) are decelerating models throughout the evolution. In order to meet
the observational requirement of an accelerating universe at present, it is
a physical necessity to propose a modified law for the variation of matter
density in the universe. In this regard we make an attempt with the
modification of the ansatz (\ref{IC8}) as

\begin{equation}
\rho =\frac{A}{R^{4}}(R^{2}+b+cR^{8})^{1/2}  \label{IC12}
\end{equation}%
where $c$ is a positive constant. The ansatz (\ref{IC12}) can alternatively
(for mathematical ease) be written as

\begin{equation}
\rho =\sqrt{\frac{\alpha }{R^{8}}+\frac{\beta }{R^{6}}+\gamma }  \label{IC14}
\end{equation}%
where $\alpha ,\ \beta ,\ $and$\ \gamma $ are all positive constants. From
the conservation equation (\ref{IC6}) \newline
for $\rho =\sqrt{\frac{\alpha }{R^{8}}+\frac{\beta }{R^{6}}+\gamma }$, the
expression for pressure $p$ is obtained as

\begin{equation}
p=\left( \frac{\alpha }{3R^{8}}-\gamma \right) \frac{1}{\sqrt{\frac{\alpha }{%
R^{8}}+\frac{\beta }{R^{6}}+\gamma }}  \label{IC15}
\end{equation}%
so that the equation of state is given by

\begin{equation}
p=\left( \frac{\alpha }{3R^{8}}-\gamma \right) \frac{1}{\rho }  \label{IC16}
\end{equation}%
which corresponds to a variable Chaplygin gas \cite{VCGi}-\cite{VCGiii}. In
view of equations (\ref{IC14}) and (\ref{IC15}), we see that 
\begin{equation}
(\rho +3p)=\frac{\left( 2\frac{\alpha }{R^{8}}+\frac{\beta }{R^{6}}-2\gamma
\right) }{\sqrt{\frac{\alpha }{R^{8}}+\frac{\beta }{R^{6}}+\gamma }}
\label{IC16A}
\end{equation}%
which is positive for very small values of the scale factor $R$ indicating $%
\ddot{R}<0$ initially (from the Raychaudhuri equation (\ref{IC6A})). The
model starts from\ the big bang with $\dot{R}\rightarrow \infty $ (from the
Friedmann equation (\ref{IC4})).

For small values of $R$, after the big bang, that is, so long as $\frac{%
\alpha }{R^{8}}\gg \frac{\beta }{R^{6}}+\gamma $, we have $\rho \approx 
\frac{\sqrt{\alpha }}{R^{4}}$ and $p\approx \frac{1}{3}\frac{\sqrt{\alpha }}{%
R^{4}}$, which corresponds to the initial radiation dominated era ($p=\frac{1%
}{3}\rho $) of the standard model. For the small values of $R$ satisfying
the condition $\frac{\alpha }{R^{8}}+\gamma \gg \frac{\beta }{R^{6}}$, we
get $\rho \approx \sqrt{\frac{\alpha }{R^{8}}+\gamma }$ and $p\approx \frac{1%
}{3}\rho -\frac{4}{3}\frac{\gamma }{\rho }$, which corresponds to a modified
generalized Chaplygin gas characterized by an equation of state $p=A\rho -%
\frac{B}{\rho ^{n}}$ \cite{MGCGi}-\cite{MGCGii}. As $R$ further increases,
the term $\frac{\beta }{R^{6}}$ starts dominating over the term $\frac{%
\alpha }{R^{8}}$ in the expression for density (\ref{IC14}). That is, the
expression for the density takes the form

\begin{equation}
\rho \approx \sqrt{\frac{\beta }{R^{6}}+\gamma }\ \ \ ,\ \ \ \left( \frac{%
\alpha }{R^{8}}\ll \frac{\beta }{R^{6}}+\gamma \right)  \label{IC17}
\end{equation}%
and the equation of state reduces to

\begin{equation}
p\approx -\frac{\gamma }{\rho }  \label{IC18}
\end{equation}%
which corresponds to a Chaplygin gas \cite{DE11}. For the values of $R$
satisfying the condition $R^{6}\ll \frac{\beta }{\gamma }$ the expression
for the density (\ref{IC17}) approximates to

\begin{equation}
\rho \approx \frac{\sqrt{\beta }}{R^{3}}  \label{IC19}
\end{equation}%
which corresponds to a FRW universe dominated by dust like matter ($p=0$).
In view of (\ref{IC17}) and (\ref{IC18}), it follows that the universe turns
from decelerated phase of expansion to one of acceleration as soon as $R^{6}>%
\frac{\beta }{2\gamma }$.

For very large values of the cosmological radius $R$ from equation (\ref%
{IC17}), the expression for the density approximates to

\begin{equation}
\rho \approx \sqrt{\gamma }  \label{IC20}
\end{equation}%
and from equation of state (\ref{IC18})

\begin{equation}
p\approx -\sqrt{\gamma }.  \label{IC21}
\end{equation}%
From Eqs. (\ref{IC20}) and (\ref{IC21}), we have $\left( \rho +3p\right) =-2%
\sqrt{\gamma }<0$ i.e. $\ddot{R}>0$ (from the Raychaudhuri equation (\ref%
{IC6A})) leading to an accelerating universe. This corresponds to an empty
universe with a cosmological constant $\sqrt{\gamma }$ (de Sitter universe).
Thus we find that the modified ansatz (\ref{IC14}) ultimately leads to an
accelerated phase of expansion passing through the different phases of
decelerated expansion, in agreement with the observations.

It is difficult to integrate the Friedmann equation (\ref{IC4}) for $\rho =%
\sqrt{\frac{\alpha }{R^{8}}+\frac{\beta }{R^{6}}+\gamma }$, even in the case 
$k=0$, to obtain the time variation of the scale factor $R$; However, the
solution for $\rho =\sqrt{\frac{\beta }{R^{6}}+\gamma }$ has already been
obtained by Kamenshchik et al. \cite{DE11}. Further detailed properties and
consequences of Chaplygin gas cosmological models have already been
discussed by Gorini et al. \cite{GOR}.

\end{document}